\documentstyle[11pt,aaspp4]{article}

\slugcomment{To appear in A.J., June 1977}

\lefthead{McCullough}
\righthead{H$\alpha$ -- Infrared Cirrus Correlation}

\begin{document}

\title{A Correlation between \\
Balmer H$\alpha$ Emission and Infrared Cirrus}

\author{Peter R. McCullough\altaffilmark{1}}
\affil{Astronomy Department, University of Illinois, Urbana, IL, 61801}

\altaffiltext{1}{Alfred P. Sloan Research Fellow.}

\begin{abstract}
A 13\arcdeg$\times$13\arcdeg~image of Balmer H$\alpha$\ emission at
galactic latitude -65\arcdeg\ is presented. Sensitivity is limited in part by
confusion\footnote{The H$\alpha$\ surface brightness at high Galactic latitudes
is approximately 1 Rayleigh (R), where
${\rm 1~R = 10^6/4\pi~photons~cm^{-2}~s^{-1}~sr^{-1}}$ (\cite{rey92}).
At H$\alpha$\, 1 R = $2.42 \times 10^{-7}$ ergs cm$^{-2}$ s$^{-1}$ sr$^{-1}$,
or an emission measure EM = 2.75 cm$^{-6}$pc for gas at 10$^4$~K.}:
the peak-to-valley anisotropy
of the H$\alpha$ surface brightness is $\sim$0.2 Rayleighs
on angular scales of 0.1\arcdeg\ to 1.0\arcdeg.
The morphology of the H$\alpha$ emission is similar to that of the
100 $\mu$m emission previously observed by the Infrared Astronomical
Satellite (IRAS).
A point-by-point comparison of the two shows a marginally
detected ($3\sigma$) positive correlation, $\rho = +0.14{\pm}0.04$.
The slope of the correlated components of the 100 $\mu$m and H$\alpha$ 
emissions is $1.26^{+0.45}_{-0.32}$ ${\rm MJy~sr^{-1}~R^{-1}}$.
Using parameters from the
literature, we predict that emission from dust and ionized gas at
high latitudes produces $\sim$3 times more H$\alpha$\ emission per unit
100$\mu$m emission than does backscattering of H$\alpha$ emission
from Galactic H~II regions by dust at high latitude.
Observations of this type may allow us to distinguish between
Galactic foreground and cosmic background for both the free-free
emission and the thermal dust emission associated with the warm ionized
medium of the Milky Way.
\end{abstract}


\section{Introduction}

Martin Harwit (\cite{har81}) might say we have rediscovered in H$\alpha$\
the interstellar cirrus that was discovered first in the infrared by IRAS
(\cite{low84}). Such a rediscovery is not insignificant.

To measure anisotropy of the cosmic microwave background, astronomers
must first measure and subtract foreground emission from the Milky Way's
interstellar medium (ISM). Three components of foreground emission from
the ISM are important (\cite{ben92}):
1) bremsstrahlung (or ``free-free'') emission from free electrons
accelerated by interactions with ions in a plasma,
2) synchrotron emission from relativistic electrons accelerated by
magnetic fields,
and
3) thermal emission from dust.
The H$\alpha$\ images directly trace the free-free component, because
ionized gas emits H$\alpha$\ and bremsstrahlung in direct proportion.
In addition, the H$\alpha$\ images indirectly trace the thermal
emission from that dust associated with the ionized medium.

This paper shows that H$\alpha$ and 100 $\mu$m emissions are weakly
correlated at high galactic latitude.  High latitude, filamentary
nebulosities have been observed before (\cite{gut94}, \cite{san76},
and references in each), but verifying the emission to be H$\alpha$ in
particular had not been possible. In fact, Guhathakurta \& Tyson (1989) showed
that the optically broad-band red color of cirrus is not due to H$\alpha$
emission, but may be due to luminescence of hydrocarbons.  Kogut et al
(1996) have shown that multi-frequency radio emissions (commensurate
with free-free) are correlated with 140 $\mu$m emission. Broadbent,
Haslam, \& Osborne (1989) have shown that free-free emission and
thermal dust emission are positively correlated at low galactic
latitude, i.e. emissions from H II regions.

At low galactic latitudes, there is considerable confusion for optical
tracers of gas: where there is gas, there tends to be dust too, and the
dust that is emitting at 100 $\mu$m, for example, is absorbing
H$\alpha$\ light.  Even at high latitudes, where dust opacity is minor
(\cite{rey80}), for any integral along a line of sight, the 100 $\mu$m
surface brightness is proportional to the density of dust, whereas the
H$\alpha$\ surface brightness is proportional to the electron density
{\it squared} (see Section \ref{discussion}).  For these reasons, we do
not expect a strict proportionality between the H$\alpha$\ and infrared
emissions.

Our H$\alpha$\ images provide not absolute surface brightnesses but
only relative ones. The same is true of the IRAS images.  We do not
attempt absolute calibration of the H$\alpha$\ surface brightness
because our H$\alpha$\ filter's 1 nm bandpass is too wide to suppress
geocoronal H$\alpha$, which is typically a few Rayleighs and varies
with time (\cite{bro68}).  Fabry-Perot spectroscopy allows absolute
calibration of H$\alpha$\ surface brightness, and the infrared has been
calibrated by COBE. However, to compare IRAS and H$\alpha$\ images, it
suffices to filter out the background in the same way for both images,
and that we have done (see Section \ref{data}).

\section{Instrumentation and Observations} \label{obs}

The observations were made at the Mt. Laguna Observatory with a
wide-field, narrowband imaging camera similar and yet superior to other
such cameras (e.g. \cite{bot88}). We attached it ``piggy-back'' upon a
1-m telescope, which was used solely for blind pointing and tracking;
no acquisition or guiding was necessary. The camera consists of a
filter wheel, a 135-mm f/2.8 lens, and a 2048$\times$2048 pixel CCD.
With the 135-mm lens, the 15-$\mu$m pixel spacing corresponds to
0.38\arcmin/pixel, and a field of view of 13\arcdeg$\times$13\arcdeg.

To tune the center wavelength of the 1.0 nm wide (FWHM)
H$\alpha$\ filter, the filter wheel was thermoelectrically cooled and
thermostatically controlled to a precision of 1\arcdeg~C. The
temperature was set such that the H$\alpha$\ filter's transmission of
zero-velocity H$\alpha$\ peaked off the optical axis while maintaining
the on-axis transmission to greater than 75\% of the peak. At off-axis
angles of $>4.8$\arcdeg, the transmission of zero-velocity
H$\alpha$\ drops below 50\%, so each image is cropped at a radius of
4.8\arcdeg.  Other than cropping, we made no attempt to correct for the
radial dependence of the sensitivity to zero-velocity
H$\alpha$\ .Considering that and other potential errors, we estimate
our absolute scaling of H$\alpha$\ surface brightness could be in error
by as much as a factor of 1.5.  Observations of external galaxies show
the velocity coverage extends from $-250$ km/s to $+250$ km/s.

The backside illuminated CCD has a quantum efficiency at 656.3 nm of
95\% (\cite{lea96}). The sensor is cooled with liquid nitrogen to
-120\arcdeg~C, so for these observations, dark current is negligible.
The noise of the readout amplifier is 9~e$^-$ rms, but the integration
times were selected so that the readout noise would be negligible
compared to the Poisson noise of the sky's photons.

Forty-eight pairs of exposures were taken over a period of 5 nights,
September 21-25, 1995. Each pair consisted of a continuum image
followed by an H$\alpha$\ image while the telescope turned at the sidereal
rate for the duration of the pair. Each continuum image was a 1-minute
exposure through a R band filter, except for the first few, which were
2 minutes each (and which had too many saturated stars). Each H$\alpha$\
exposure was 20 minutes; the total observing time on-source was 16 hours.
The images were dithered; that is, the
telescope was repositioned after each pair of images. The dithering
pattern was a 5$\times$5 grid with 1\arcdeg\ spacing, covered almost
twice with the 48 image pairs. 

We observed the North America nebula and Barnard's Loop to calibrated the
H$\alpha$\ surface brightness. These two H~II regions have surface brightnesses
of 850 R and 153 R (each $\pm$10\%, \cite{rey80}),
averaged over 48\arcmin\ diameter fields
centered on 21h00m.0 +45\arcdeg 00\arcmin\ [2000] and
05h53m.8 -06\arcdeg 08\arcmin\ [2000].

\section{Data Reduction} \label{data}

Much of the data reduction is aimed at eliminating stars from the
images while preserving the diffuse H$\alpha$\ emission.
The stars are suppressed in a number of ways:  1) continuum
subtraction, 2) flagging pixels that are bright in continuum, 3)
flagging pixels at the locations of filter-induced flares near very
bright stars, and 4) median filtering with a spatial scale larger than
a stellar image.

The CCD bias was determined by overscanning.  The relative gain
correction for the CCD pixels (the ``flat field'') was determined by
observing the zenith at sunset in order to minimize gradients and
curvatures (\cite{chr96}).  Each continuum image was scaled by a
constant and subtracted from the H$\alpha$\ images.  The constant was
empirically-determined based upon stars in each field.

The continuum-subtracted images invariably exhibited a smooth intensity
pattern with radial symmetry about the optical axis. The radial pattern
may be due to telluric line emission, although it
could also be due to the combination of vignetting and inaccuracy of the
flat fields. In any event, the circular pattern was removed by
fitting and subtracting circular isophotes.

Our experience has been that continuum subtraction is adequate for
faint stars, but inadequate for bright stars. Invariably, the two images
are not focused identically or they are not registered well enough or both.
In the presence of these systematic errors and the increased
Poisson noise at the positions of bright stars, it is worthwhile to excise
those pixels near bright stars from the data. We flag pixels in the H$\alpha$\
images based upon their respective intensities in the R band images,
and flagged pixels are excluded when computing the medians
described below.

Each image was median filtered spatially with a
2\arcdeg\ diameter ring median filter (\cite{sec95}). Each median-filtered
image was subtracted from its original to eliminate large scale
intensity variation due to the variation in airmass across
the field, residual inaccuracies in flat fielding, or the occasional
telluric cirrus cloud. To check that this procedure was not eliminating
real emission from the Milky Way, we also reduced the
data without this median filtering step, but instead substituted fitting and
subtracting a plane of intensity. The two methods give similar results.

The images were registered automatically to a common coordinate system, using
a few dozen SAO stars on each image to
determine the appropriate shift, expansion, rotation, and distortion.
The 48 registered images were combined by averaging them pixel by pixel,
while excluding flagged pixels. Also, an iterative procedure eliminated those
individual pixel values that were further than $2\sigma$ from the median of the
other values for that particular pixel. Finally, a ring median filter of
radius 2.3\arcmin\ was applied, and then the image was convolved with a
circular Gaussian of width $\sigma = 1.5$\arcmin.

\section{Results} \label{results}

\begin{figure}
\plotone{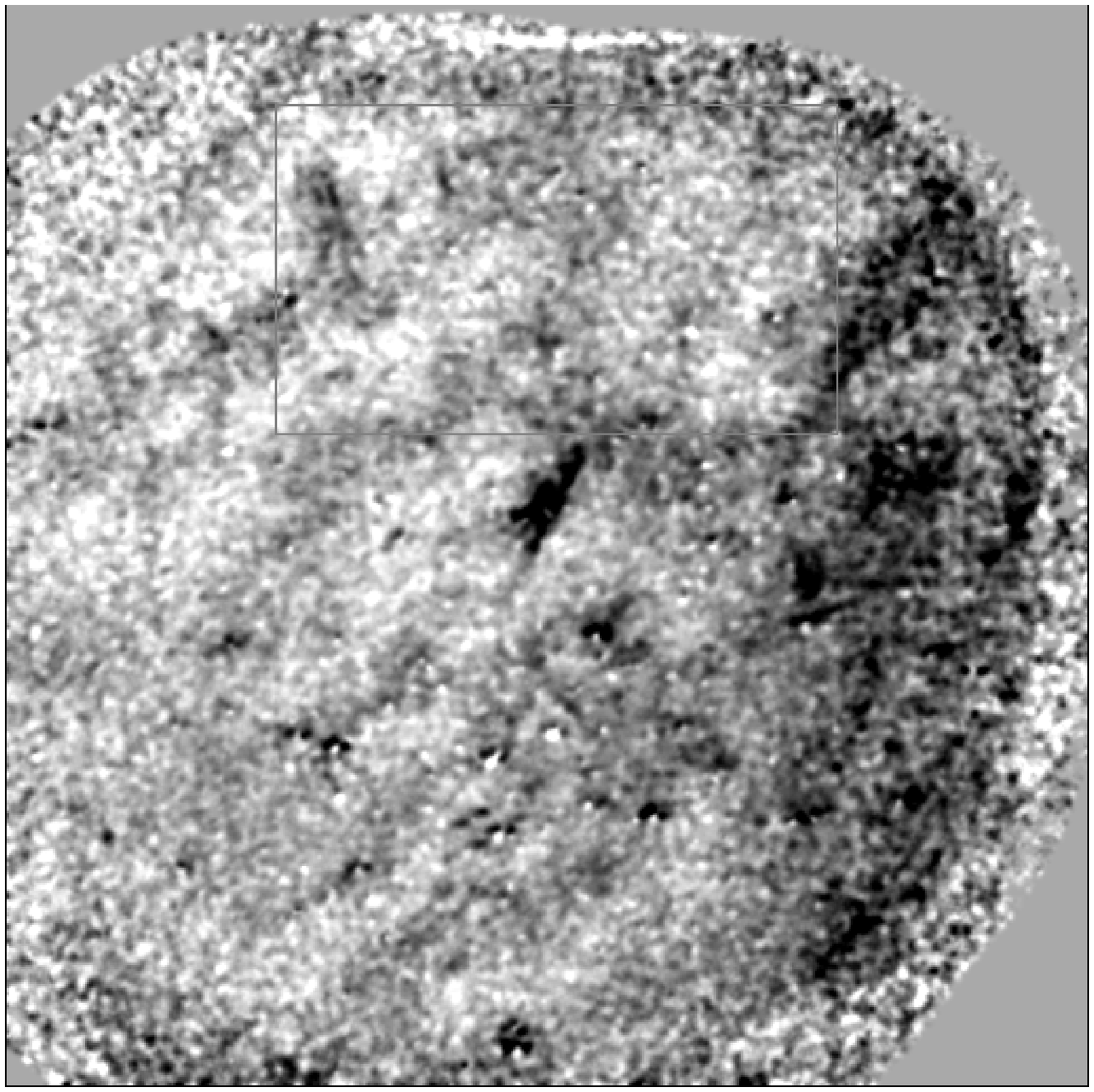}
\caption{
This H$\alpha$\ image is 13\arcdeg\ on a side and shows features as faint as
0.2 Rayleighs. The triangular object near the center is the
brightest object in the field, at 0.6 R.  The inscribed box contains
two H$\alpha$\ emission features associated with IRAS 100$\mu$m emission:
the arc at left and the potato-shaped outline at top-center (cf. Figure
2).
The center of the image is
23h40m00s -12\arcdeg 00\arcmin 00\arcsec [2000],
(l,b) = (71\arcdeg .5,$-67$\arcdeg .4). 
North is up and East is to the left.
The Galactic plane is toward the upper-right (NW).
\label{fig1}}
\end{figure}

\begin{figure}
\plotone{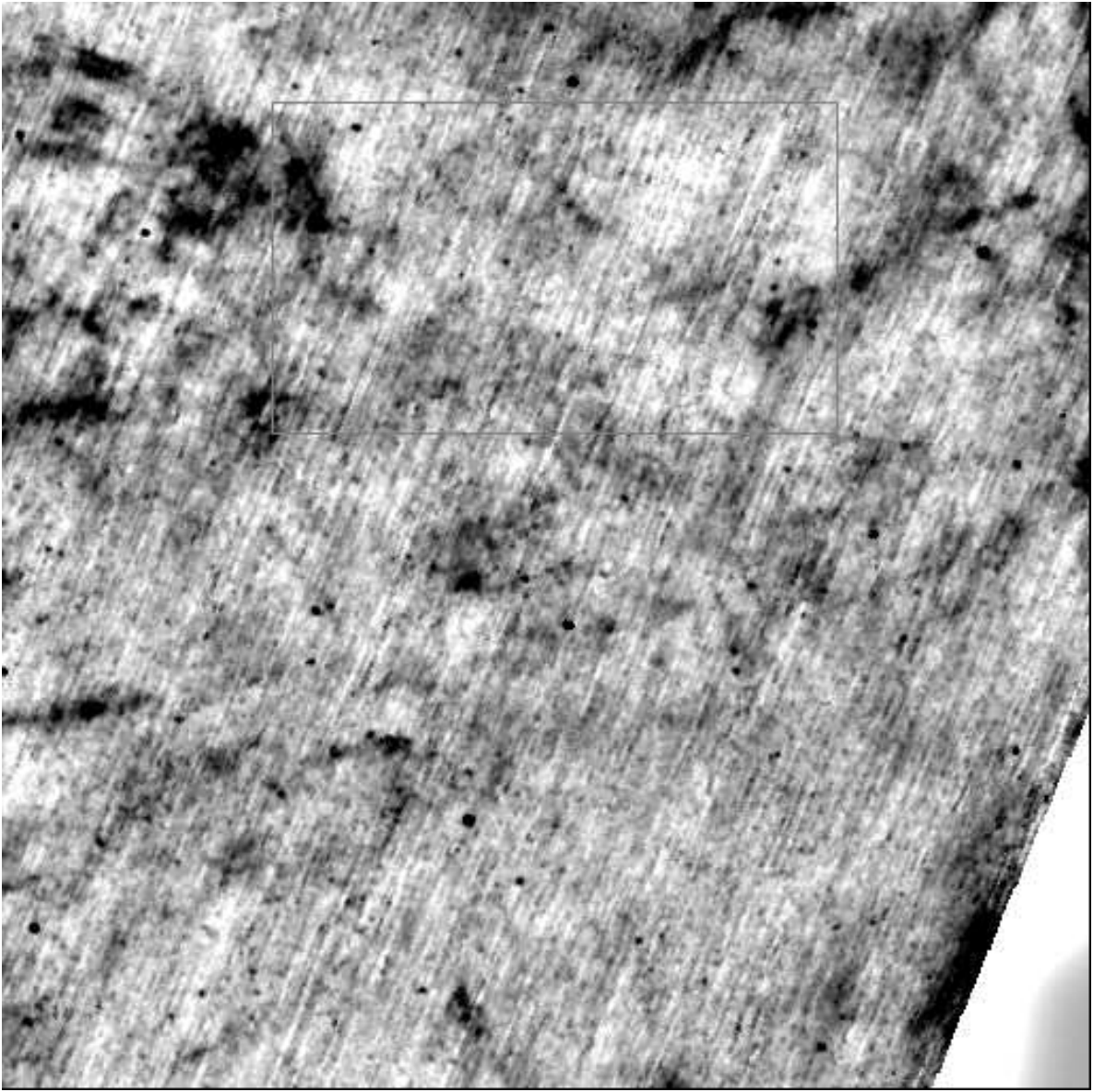}
\caption{
This IRAS 100$\mu$m image shows the same region as Figure 1.
Because of its proximity to the ecliptic, this field is one of the
so-called ``reject'' fields of IRAS.
To reduce variations on large angular scales, a 2\arcdeg\ diameter
ring-median filtered version of this image was subtracted from
the original. The same filtering was applied to Figure 1.
\label{fig2}}
\end{figure}

The H$\alpha$\ image (Figure \ref{fig1}) appears similar to the
corresponding area of sky observed by IRAS at 100 $\mu$m (Figure
\ref{fig2}).  We have inscribed boxes in Figures \ref{fig1} and
\ref{fig2} that contain the 
two most apparent similarities: 1) at left, a
filled arc of 100 $\mu$m emission, enveloped by an arc of H$\alpha$\
emission, and 2) at center-top, a potato-shaped outline in both 100
$\mu$m and H$\alpha$\ emission.\footnote{A ``movie'' that ``blinks'' the
two H$\alpha$\ and 100 $\mu$m images is available from the author.}

To quantify the association between the H$\alpha$ and 100 $\mu$m
emissions, we evaluated their (Pearson) correlation coefficient, $\rho =
+0.115$, over the central
8.5\arcdeg$\times$8.5\arcdeg\ subregions of the
13\arcdeg$\times$13\arcdeg\ images. The sugregion was selected
because the effective observing time is less near all edges of
the H$\alpha$\ mosaic and because residual, circular artifacts
of the mosaicing process are apparent near the western edge.

\begin{figure}
\plotone{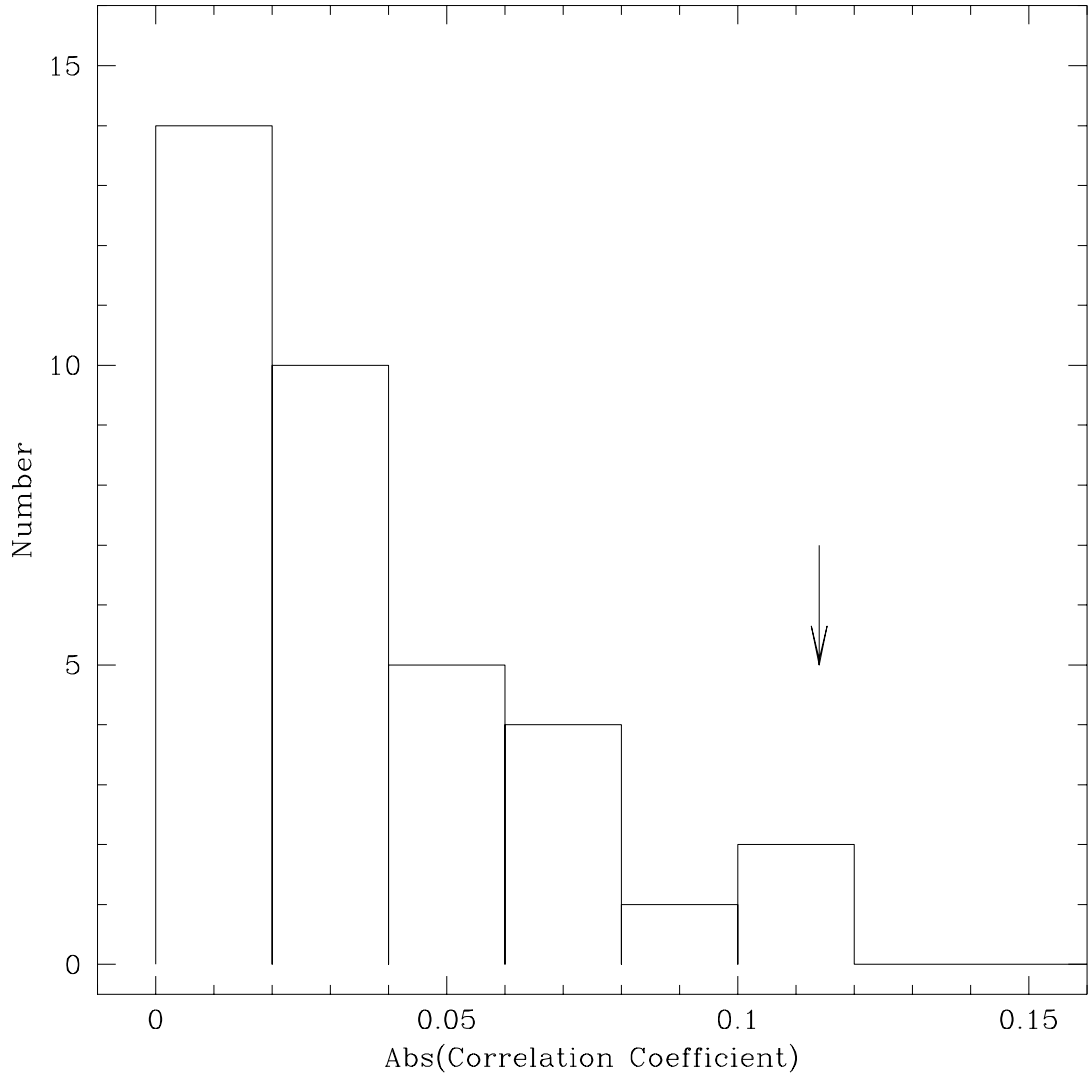}
\caption{
This histogram of the absolute values of the correlation
coefficients of 36 controls shows that the correlation coefficient
of the H$\alpha$\ and 100$\mu$m emissions in Figures 1 and 2, $+$0.115 (arrow),
is statistically significant at 3$\sigma$.
\label{fig3}}
\end{figure}

We also correlated the H$\alpha$ image with 36 controls.
The controls are 100
$\mu$m IRAS images from ``random'' but statistically similar
directions on the sky, centered at galactic coordinates b =
${\pm}67$\arcdeg , l = 1\arcdeg , 21\arcdeg ,
..., 341\arcdeg .
Each control was created and processed like the image in Figure 2.
The controls should be uncorrelated with the
H$\alpha$ image (i.e. the expectation value, ${\rm E(\rho) = 0}$),
so they provide a standard by which to judge the validity of our measured
correlation. The distribution of
correlation coefficients of the 36 controls is approximately
Gaussian, with $\mu$ = -0.017 and $\sigma$ = 0.041 (see Figure
\ref{fig3}). 

The measured correlation coefficient ($\rho = +0.115$)
is statistically significant from zero by $+2.8\sigma$.
From the controls we can conclude with $\sim$95\% confidence\footnote{That 2 of
36 controls share the same bin in Figure 3 as the true correlation
coefficient indicates a $\sim$5\% likelihood that the two emissions
are uncorrelated ($\rho = 0$), i.e. that the measured correlation
is due solely to chance.}
that $\rho \not= 0$, but we should not immediately conclude
that $\rho = +0.115{\pm}0.041$.
Mathematically, we expect $\rho$ to be larger than $+0.115$ because 
noise in the data tends to reduce $|{\rho}|$ from its true value. 
Physically, we expect $\rho$ to be larger or smaller for various directions
on the sky. For example, in the boxes of Figures 1 and 2, which were
inscribed because the emissions within them looked similar, $\rho = +0.150$.

The bias of $\rho$ due to noise in the two images can be
estimated, and it is not large. We let ${\rm \sigma_i^2}$ equal the 
variance of the infrared intensity and we let ${\rm \sigma_h^2}$ equal the
variance of the H$\alpha$ intensity, both as measured over the inner
8.5\arcdeg$\times$8.5\arcdeg\ subregions. The ${\rm \sigma_i^2}$ and
${\rm \sigma_h^2}$ each include both genuine variation of emission and 
random noise.  We evaluate
the noise variances in each image, ${\rm \eta_i^2}$ and ${\rm \eta_h^2}$,
by computing the average of the variances in small regions apparently
lacking emission. If the ratios
${\rm \sigma_i^2/\eta_i^2}$ and ${\rm \sigma_h^2/\eta_h^2}$ are both large
compared to unity, the bias is small and more precise estimates of the
noise are not warranted.
The measured values are ${\rm \sigma_i/\eta_i = 2.4}$ and
${\rm \sigma_h/\eta_h = 2.6}$.
In can be shown that the ratio of the true to
the observed (Pearson) correlation coefficients is 
\begin{equation}
{\rm {{\rho_{true}}\over{\rho_{obs}}} =
\sqrt{ {{\sigma_i^2}\over{\sigma_i^2-\eta_i^2}} }
\sqrt{ {{\sigma_h^2}\over{\sigma_h^2-\eta_h^2}} },}
\label{rhoratio}
\end{equation}
which in this case is 1.19 and is smaller than the fractional
uncertainties of both the statistical variation
of $\rho$ estimated from the controls and the spatial variation evident
from the value of $\rho = +0.150$ of the alternative subregion.
Accounting for the bias of noise, we estimate $\rho = +0.14{\pm}0.04$.

Having established that the 100 $\mu$m and H$\alpha$ emissions
are weakly correlated, we evaluate the ratio of the two correlated
emissions by a linear fit of the form,
I(100 $\mu$m) = a I(H$\alpha$) + b, to the data within the
central 8.5\arcdeg$\times$8.5\arcdeg\ subregions of the two images.
The slope ``a'' is of interest; the value ``b'' is not, 
because the zero points of both the 100 $\mu$m intensity and the H$\alpha$\
intensity are arbitrary.
Because both the 100 $\mu$m and H$\alpha$ emissions have substantial
observational errors, ordinary least squares fitting is invalid.\footnote{In
this context ``ordinary'' least squares refers to fitting a ``dependent''
(noisy) variable Y against an ``independent'' (noise-free) variable X.}
We fit the data by two methods as described below.
Both methods allow for errors in each of the observed quantities.

The first method is based upon the work of
\cite{fei92}.\footnote{Note that their variable $l = \sigma_y^2 /
\sigma_x^2$, not $\sigma_x^2 / \sigma_y^2$ as originally printed.} It
yields a value for the slope a = 1.2(${\pm}0.12$) ${\rm MJy~sr^{-1}~R^{-1}}$,
under the assumption of additive Gaussian errors of zero mean and variances of
(0.17 MJy sr$^{-1}$)$^2$ and (0.077 R)$^2$, for the 100 $\mu$m and
H$\alpha$\ emissions, respectively.
We determined the variance of the
H$\alpha$\ emission, (0.077 R)$^2$, from an ensemble of 1-dimensional
Gaussian fits to the bivariate distribution of 100 $\mu$m and H$\alpha$
emissions by rows; the variance of the 100 $\mu$m emission was
determined similarly, but by columns (cf. Figure 4).
Note that those variances include
real (but uncorrelated) emission in addition to detector and shot noise;
the uncorrelated emissions act as approximately Gaussian ``errors''
with respect to the correlated components of the emissions.
We consider the formal error (${\rm \sigma_a = 0.12 MJy~sr^{-1}~R^{-1}}$)
to be a lower limit of the true error.
To investigate empirically the reliability of the slope
estimate, we performed the same analysis separately on the four quadrants
of the central 8.5\arcdeg$\times$8.5\arcdeg\ subregion.
The slopes so measured were 1.5, 1.4, 1.3, and 0.6
${\rm MJy~sr^{-1}~R^{-1}}$ for the NW, NE, SW, and SE quadrants, respectively.

The second method is a simplification of the ``BCES'' method of
\cite{akr96}. BCES refers to ``bivariate correlated errors and
intrinsic scatter''; our simplification is to assume that the
observational errors of the 100 $\mu$m and H$\alpha$ emissions are
uncorrelated. We evaluate the observational errors by examining small
regions apparently lacking emission. The standard deviations so
measured are 0.07 MJy~sr$^{-1}$ and 0.04 R, for the 100 $\mu$m and H$\alpha$
emissions respectively. The BCES bisector slope (see \cite{akr96}),
evaluated over the central 8.5\arcdeg$\times$8.5\arcdeg\ subregion,
yields the same slope as the first fitting method,
a = 1.2 ${\rm MJy~sr^{-1}~R^{-1}}$.
Likewise, the BCES bisector slopes for the NW, NE, SW, and SE quadrants
are 1.68, 1.31, 1.16, and 1.53 ${\rm MJy~sr^{-1}~R^{-1}}$, respectively.

The result of averaging the 8 slopes from the four quadrants and the
two methods is a = $1.26^{+0.45}_{-0.32}$  ${\rm MJy~sr^{-1}~R^{-1}}$.
The mean value given is
a vector average (i.e. ${\rm <a>~=~tan(<arctan(a_i)>)}$, i = 1, ..., 8).
The excursions ($+0.45$ and $-0.32$) describe the 1$\sigma$ dispersion
of the 8 slopes, also calculated vectorially, not arithmetically.
Additional observations are required to determine
if the dispersion of slopes is statistical or intrinsic to the ISM.

The brightest object in the H$\alpha$\ image has no counterpart at 100
$\mu$m, illustrating that the correlation between 100 $\mu$m and
H$\alpha$\ emissions is not true in general.  To the contrary,
additional observations at lower galactic latitudes ($|b| < 30$\arcdeg)
and of less sensitivity show many H$\alpha$\ features with no infrared
counterparts (\cite{mcc97}).  Those additional images also show
H$\alpha$--infrared correlations, both positive and negative.  At the
lower Galactic latitudes, there is considerable confusion.

The brightest object in Figure \ref{fig1} may be associated
with the Magellanic Stream; it lies between two H~I peaks of MS IV
(\cite{coh82}) and its long axis is parallel to the Magellanic Stream.
If so, it is the brightest H$\alpha$\ emission from the Magellanic Stream
(cf. \cite{waw96,wei96}). However, it might be simply a density enhancement in
the ISM unrelated to the Magellanic Stream.
The critical test is to measure its velocity.

\section{Discussion} \label{discussion}

The H$\alpha$\ light we observe could be from
either a reflection nebula or an emission nebula, or both.
In this Section we show that the reflection component is expected to be
small compared to the reflection component.

In the case of a reflection nebula, the surface brightness of the H$\alpha$\
light is proportional to the product of the surface emissivity in H$\alpha$\
of the disk of the Milky Way and the dust column density along the line of
sight. Using canonical values for the surface emissivity, the ratio of dust to
gas, and the dust's scattering cross section and phase function,
Jura (1979) predicts the surface brightness of the H$\alpha$\ 
light I$_\alpha$ (in Rayleighs) backscattered from dust:
\begin{equation}
{\rm I_\alpha = 0.10 {{N_H}\over{10^{20} cm^{-2}}},}
\end{equation}
where ${\rm N_H}$ is the total hydrogen column density.

In the case of an emission nebula, the surface brightness of the H$\alpha$\
light I$_\alpha$ (in Rayleighs) is proportional to the emission measure (\cite{rey92}):
\begin{equation}
{\rm I_\alpha = 0.36 T_4^{-0.92} EM,}
\label{recomb}
\end{equation}
where T$_4$ is the electron temperature in units of $10^4$~K. In the warm
ionized medium of the ISM, T$_4 \approx 0.8$ (\cite{rey92}).
The emission measure
EM $={\rm \int n_e n_p dl \approx \int n_e^2 dl}$ has units
cm$^{-6}-$pc. The region is assumed to be optically thick to Lyman lines
(a.k.a. Case B), and extinction is ignored, but otherwise, equation \ref{recomb}
is derived directly from the fundamental physics of recombination.

From the emission measures and dispersion measures along the high-latitude lines
of sight to a few globular clusters containing pulsars, Reynolds (1991a)
shows that the density of high latitude ionized gas,
where it exists\footnote{Reynolds estimates that the ionized gas fills
${\sim}20$\% of the volume.}
is $\sim$0.08 cm$^{-3}$ and the ratio of the ionized hydrogen
column density to the total hydrogen column density ranges from 0.2 to
0.4 (\cite{rey91b}), with an average of 0.27 for the five lines of sight.

Combining the emission and reflection contributions to the H$\alpha$\
surface brightness using the above values from the literature, we find
\begin{eqnarray}
{\rm I_\alpha} & = & {\rm (11.7 T_4^{-0.92} n_e {{N_{H^+}}\over{N_H}} + 0.10) {{N_H}\over{10^{20} cm^{-2}}},} \nonumber \\
& = & {\rm (0.31 + 0.10) {{N_H}\over{10^{20} cm^{-2}}},}
\label{Iha}
\end{eqnarray}
where we have expressed the EM as a product of the
volume density, n$_e$, and the ionized hydrogen column density, N$_{H^+}$,
by approximating 
${\rm \int n_e n_p dl \approx n_e \int n_{H^+} dl}$.
The latter approximation is valid if hydrogen is the dominant ion and if
the electron volume density is
bimodal, i.e. either ${\rm n_e}$ is
0.08 cm$^{-3}$ or it is zero.

Equation \ref{Iha} predicts the emission component to be $\sim$3 times
greater than the backscattered component. The backscattering component
will be less than a third of the emission component if the gas is either
denser or more highly ionized than we have assumed.
Reynolds et al (1973) estimate 
the ratio of backscattered to total H$\alpha$\ light to be 0.1 ($-0.1$,$+0.2$)
toward a few Lynds nebulae.

It is well known that dust and neutral hydrogen are positively correlated
in general (\cite{bou88}), but is that true in this region?
The three H~I maps of highest resolution and sensitivity
(\cite{har94},\cite{hei74},\cite{coh82}) are inadequate to allow us to
answer this question because the beams are too large (0.5\arcdeg\ FWHM)
and the sensitivity is too poor (the r.m.s. is equivalent to 
${\rm N_H} \approx 1$ to a few times 10$^{19}$ cm$^{-2}$).
The {\it peak} 100 $\mu$m emission,
$\sim$0.5 MJy sr$^{-1}$ above the background,
corresponds to a {\it peak} H~I column density of
6$\times$10$^{19}$ cm$^{-2}$,
assuming ${\rm I_{100} = 0.85 {{N_H}/{10^{20} cm^{-2}}}}$ MJy sr$^{-1}$
(\cite{bou88}).
Nevertheless, we analyzed the H~I column density data of 
Heiles \& Habing (1974) exactly as we did
the H$\alpha$ data above and found no statistically significant
correlation with the
100 $\mu$m data.\footnote{The data of \cite{coh82} exist only on 8-hole 
paper tape, and the data of \cite{har94} were not yet publicly 
available in machine readable form when this paper was written.}
The H~I\&infrared correlation coefficient is $-0.119$, which 
corresponds to $1.5\sigma$ less than zero
($\sigma = 0.078$ is evaluated as before using thirty six 100 $\mu$m
controls). The anticorrelation of H~I\&infrared in the same region
where H$\alpha$\&infrared is positively correlated 
is intriguing, but it is not statistically significant.
The anticorrelation also is opposite in sign to the firmly 
established results of Boulanger \& Perault (1988).

In Section \ref{results} the slope of the correlated
components of the 100 $\mu$m and H$\alpha$ emissions is estimated to be
$1.26^{+0.45}_{-0.32}$  ${\rm MJy~sr^{-1}~R^{-1}}$.
From 4 years of COBE data, Kogut et al. (1996) determine the slope of
the correlated components of the antenna temperature ${\rm T_A}$ of the
DMR at 31.5 GHz with respect to the 140 $\mu$m surface brightness ${\rm
I_{140}}$ measured with DIRBE:  ${\rm T_A / I_{140} = 6.37({\pm}1.52)~{\mu}K
(MJy~sr^{-1})^{-1} }$.  In so far as free-free emission
dominates thermal dust emission at 31.5 GHz, the slope ${\rm
6.37({\pm}1.52)~{\mu}K~(MJy~sr^{-1})^{-1}}$ corresponds\footnote{See
e.g. \cite{ben92} equation 4.} to a ratio ${\rm I_{140} / I_{H\alpha} =
0.90^{+0.29}_{-0.17}~MJy~sr^{-1}~R^{-1}}$,
which is similar to 
the ${\rm I_{100} / I_{H\alpha}}$ ratio derived in this work. 
On the $\sim$7\arcdeg\ angular scale of DMR, \cite{kog96} conclude that more
than one third of (and perhaps ``the bulk of'') the free-free emission
must be from that component correlated with 140 $\mu$m emission. 
At much smaller angular scales, the bulk of the
H$\alpha$ emission is not correlated with 100 $\mu$m emission
(Section \ref{results}); thus it
appears that the correlation between ionized gas and dust
emissions increases with increasing angular scale.

\section{Summary}

An H$\alpha$\ image of a 13\arcdeg$\times$13\arcdeg\ region at
(l,b) = (71\arcdeg,$-67$\arcdeg ) shows structure $\sim$10 times fainter
than the 12\arcdeg$\times$10\arcdeg\ region of sky at 
(l,b) = (144\arcdeg,$-21$\arcdeg ) previously observed by Reynolds (1980).
The new image gives a preliminary view of the sky over many degrees
at unprecedented sensitivity.
The new H$\alpha$\ image weakly resembles the 100 $\mu$m IRAS image;
a point-by-point comparison of the two shows a positive correlation
($\rho = +0.14\pm0.04$); and comparison with COBE data indicates that 
the correlation coefficient increases with increasing angular scale.
The slope of the correlated components of the 100 $\mu$m and H$\alpha$ 
emissions is $1.26^{+0.45}_{-0.32}$ ${\rm MJy~sr^{-1}~R^{-1}}$.

An important extension of this work (already underway by many investigators)
will be to image the H$\alpha$ emission over much of the sky. An all-sky
H$\alpha$ image may prove critical to distinguishing the Galactic
foreground from the cosmic background at the precision necessary to
determine cosmological parameters.



\acknowledgments

Jeremy Hicks assisted with the observing.
Bob Leach made the CCD camera.
We appreciate discussions with
John Gaustad,
Carl Heiles,
Margaret Meixner,
William Reach,
Ron Reynolds,
and
Dave Van Buren.
We thank Ben Weiner for providing data in advance of publication.

\clearpage

\begin{figure}
\plotone{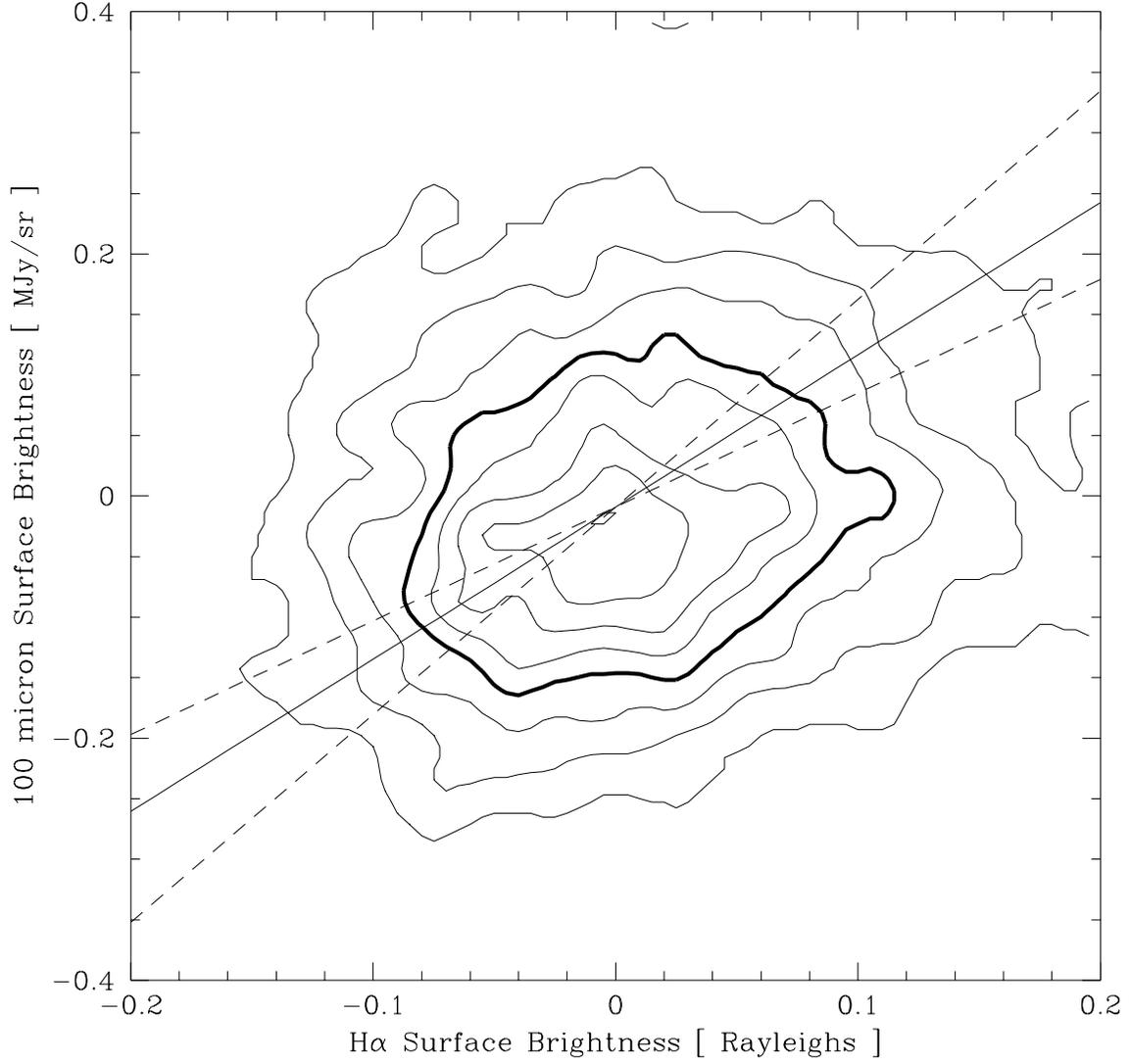}
\caption{
The 100$\mu$m versus H$\alpha$\ bivariate distribution is plotted
(contours); the best fit line (solid) and lines with ${\pm}1\sigma$
deviations of the slope (dashed) are superimposed. The contours are
linearly spaced at intervals of 12.5\%, 25\%, ... of the peak; the
50\% contour is bold.
The data are from the central 8.5\arcdeg$\times$8.5\arcdeg\ subregions of the
13\arcdeg$\times$13\arcdeg\ images in Figures 1 and 2.
\label{fig4}}
\end{figure}

\end{document}